\documentclass[aps,prl,reprint,showpacs,showkeys,groupedaddress]{revtex4-1}

\usepackage{amsmath, amssymb}
\usepackage{graphicx}
\usepackage{subfigure}
\usepackage{color}
\newcommand*{\rom}[1]{}
\begin{document}


\title{Long range interaction induced density modulated state in a Bose-Einstein condensate}

\author{Abhijit Pendse}\thanks{abhijeet.pendse@students.iiserpune.ac.in}
\author{A. Bhattacharyay}\thanks{a.bhattacharyay@iiserpune.ac.in}
\affiliation{Indian Institute of Science Education and Research}

\date{\today}

\begin{abstract}
We consider a Gross-Pitaevskii model of BEC with non-local s-wave scattering to study the density modulated state in 1D. We resort to a perturbative Taylor series expansion for the order parameter. By perturbative calculations, we show that under long range s-wave scattering a density modulated state is energetically favourable as compared to the uniform density state. We obtain density modulated state as a solution to the perturbative non-local GP equation, rather than the conventional approach of introducing amplitude modulations on top of the uniform density state and lowering the roton minimum.
\end{abstract}


\maketitle

\section{Introduction}
The Bose-Einstein condensate (BEC) is a superfluid macroscopic quantum phase of matter. It has tremendous importance in applications across a wide range of physics. To mention a few, BEC has applications in the field of Quantum Information\cite{1}, Quantum Metrology\cite{2}, Atomic Lasers\cite{3,4}, Atom Holography\cite{5}, Interferometry\cite{6}, Slow Light\cite{7}, Atom Clocks\cite{8}, Analogue Gravity\cite{9} and Quantized Vortex Dynamics\cite{sub}. Normally, the considered uniform ground state of this inherently unstable gas phase at nano-kelvin temperatures is given by the complex order parameter $\psi_0=\sqrt{n}e^{-i\mu t/\hbar}$, where, $n$ is the density of the condensate and $\mu$ is the chemical potential. In such a system, one considers two-body s-wave scattering to be the means of interaction between bosons and the s-wave scattering length $a$ to be much smaller than the average inter-particle separation $n^{-1/3}$. Small amplitude excitations of the uniform ground state determine the thermodynamics of the system and in this respect the work done by Fetter et al. is interesting\cite{fetter}. The ground state is dynamically stable to small amplitude fluctuations of the form $\theta({\bf r},t)=\sum_i{[u_i({\bf r})e^{\frac{-i\omega_i t}{\hbar}}+v^*_i({\bf r})e^{\frac{i\omega_i t}{\hbar}}]}e^{-i\mu t/\hbar}$ provided $\int{d{\bf r}|u_i|^2}\neq\int{d{\bf r}|v_i|^2} $. These small amplitude excitations are important in determining the thermodynamics of this short lived ground state of BEC.
\par 
Because of the superfluid character of the BEC, for the velocities below the velocity of sound in it, a modulated density phase is of particular interest. Supersolid is a state of matter with a crystalline order flowing without dissipation. Penrose and Onsager (PO) \cite{PO} have shown the impossibility of having such a phase (considering superfluid helium). Since then, many have contended this result and tried to circumvent the PO observations by postulating the presence of a lattice of vacancies in the solid and considering a super-flow of these vacancies \cite{ches,legg}. There are situations where there are not always particles sitting at each lattice site as has been modelled by PO. Some work in this direction on BEC have been done by considering lowering of the roton minimum\cite{PR93,PR94}. There also exist some other recently given interesting proposals based on dynamical creation of super-solid in optical lattice \cite{tass} and using Rydberg-excited BEC \cite{nath}. 
\par
The present work is motivated by the idea of obtaining a density modulated state with lower energy than that of the uniform density state. The usual approach for obtaining such a state is working in the vicinity of a uniform ground state, in other words introducing modulations on top of a uniform density state and then lowering the roton minimum. But, we take a different approach than the usual one and look for a pure density modulated state which is \textit{not} the same as modulations on top of the unifrom density state. We show that long range s-wave scattering length can indeed support such a pure density modulated state. Feshbach Resonance in BEC\cite{kett} could be of use in getting the proposed state here, since it lets us have control over the s-wave scattering length by tuning the scattering length in the interval $(-\infty,\infty)$.
\begin{figure}[h]
   \includegraphics[width=8cm]{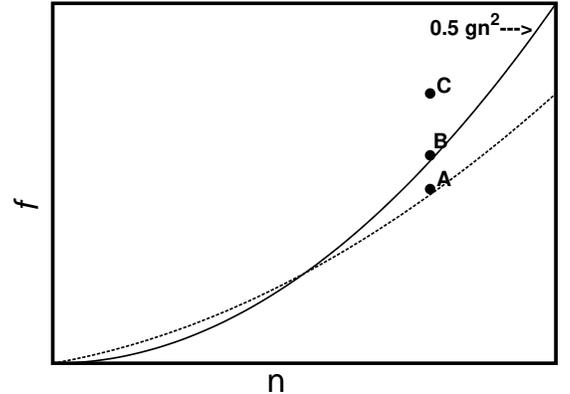}
   \caption{Figure shows a schematic diagram of free energy versus density for the uniform density state and for the desired density modulated state.}
\label{fig:schematic}
\end{figure}

\par
The zero temperature free energy density($f$) vs. density($n$) plot is shown schematically for a uniform density phase by the continuous line in Fig.\ref{fig:schematic}. The free energy density $f=\frac{1}{2}gn^{2}$ is the free energy density of the uniform BEC with contact interactions between particles where $n$ is the density of the condensate and $g$ is the s-wave interaction strength. At a density $n$, the uniform phase hangs on this energy curve, for example, at point $B$ as shown in the figure. A single particle phase characterized by a wave number(spatial order) would have a kinetic energy cost on top of this free energy density and would be somewhat at a higher energy for the same density( for example, the point $C$ in Fig.\ref{fig:schematic}). Now, in the presence of long range s-wave scattering between the particles, we are interested here in getting a free energy curve which is schematically shown by the broken line in Fig.\ref{fig:schematic}. On this, so far as the energy associated with the wave number dominates at low density, the curve remains over the $f=\frac{1}{2}gn^{2}$ curve. At a higher density where the kinetic energy cost is relatively smaller than interactions, we expect to see the curve to cross the $f=\frac{1}{2}gn^{2}$ curve and come below it such that there is now a lower energy state as shown by the point $A$ with a periodic order. This is a different approach than the conventional one aimed at lowering the roton minimum obtained on the dispersion curve in the vicinity of the uniform phase. Here, we are interested in looking directly at the role of the nonlinear interaction term in getting us to such a lower energy phase based on long range interactions. In what follows, we would see that in the presence of long range scattering, there would exist a range of boundary conditions which can lower the free energy curve as shown in Fig.\ref{fig:schematic}.

\section{The model}
The general Gross-Pitaevskii(GP) equation for a condensate is given as

\begin{equation}
\begin{split}
i\hbar\frac{\partial\psi(\bf{r,t})}{\partial t}&= -\frac{\hbar^{2}}{2m} \nabla^{2} \psi(\bf{r,t})\\
&+ \psi(\bf{r,t}) \int_{r}^\infty{\bf{dr^{'}} \psi^{*}(\bf{r',t}) V(\bf{r'-r}) \psi(\bf{r',t})} ,
\end{split}
\label{eq:original}
\end{equation}

 which can be derived from the energy functional

\begin{equation}
\begin{split}
E&=\frac{\hbar^{2}}{2m}\int{\bf{dr}|\nabla \psi(r)|^{2}}\\
&+\int{\bf{dr}\frac{|\psi(r)|^{2}}{2}\int{\bf{dr^{'}}\psi^{*}(\bf{r^{'}}) V(\bf{r-r^{'}})\psi(\bf{r^{'}})}}
\label{eq:energy}
\end{split}
\end{equation}
The local form of the GP equation considers only contact type interactions between particles. This local form of the GP equation is

\begin{equation}
i\hbar\frac{\partial\psi(\bf{r,t})}{\partial t}= -\frac{\hbar ^{2}}{2m} \nabla^{2} \psi(\bf{r,t}) + g \psi(\bf{r,t})|\psi(\bf{r,t})|^{2} .
\label{eq:contact}
\end{equation}

These equations have been successful in explaining many properties of a BEC.
\par
Let us look at features of Eq.(\ref{eq:original}) if the inter-particle interactions are taken to be non-local. Since, we are considering s-wave interactions only, we have the benefit of using any effective, soft interparticle potential $V_{eff}$ as long as it satisfies the criterion $\int V_{eff} \bf{dr^{'}}= g =\frac{4\pi\hbar^{2}a}{m}$ . We use this property and consider a potential which is a gaussian with standard deviation equal to the s-wave scattering length.
\par
Throughout the article, we use Cartesian coordinate system. Also, we will be considering here a condensate in the absence of an external potential.

\par
Let us look at a perturbative one dimensional solution of Eq.(\ref{eq:original}) by considering the interaction potential of the form $V_{eff}(\bf{r'-r})=\frac{g}{(\sqrt{2\pi}a)^{3}} \left(e^{-\frac{|\bf{x-x'}|^{2}}{2a^{2}}}\cdot e^{-\frac{|\bf{y-y'}|^{2}}{2a^{2}}}\cdot e^{-\frac{|\bf{z-z'}|^{2}}{2a^{2}}}\right)$ . By one dimensional solution, we mean that a solution of the form $\psi(\textbf{r},t)=\psi(x,t)\psi(y)\psi(z)$ where $\psi(y)$ and $\psi(z)$ are constants and only $\psi(x,t)$ varies. Using this form, we consider the integral term in Eq.(1). Since we have a symmetric potential, the terms with odd orders of derivatives will vanish due to symmetry. Also, only derivatives of $\psi(x,t)$ would be present as we have set $\psi(y)$ and $\psi(z)$ as constants. By expanding $\psi(\textbf{r}^{'},t)$ around $\textbf{r}\equiv(x,y,z)$, we get the following expression

\begin{widetext}
\begin{equation*}
\begin{split}
\psi(y)\psi(z)\left(i\hbar\frac{\partial \psi(x,t)}{\partial t}\right)&= \psi(y) \psi(z)\Big[- \frac{\hbar^{2}}{2m} \partial_{x}^{2}\psi(x,t)+\frac{g}{\sqrt{2\pi}a} \psi(x,t)\Big( |\psi(x,t)|^{2}\int_{-\infty}^{\infty}{dx' e^{-\frac{|x-x'|^{2}}{2a^{2}}}}\\
&+ (\partial_{x}^{2}|\psi(x',t)|^{2}\,) |_{\bf{x'=x}} \int_{-\infty}^{\infty}{dx' \frac{|x-x'|^{2}}{2!} e^{-\frac{|x-x'|^{2}}{2a^{2}}}}+\\
&  (\partial_{x}^{4}|\psi(x',t)|^{2}\,) |_{\bf{x'=x}} \int_{-\infty}^{\infty}{dx' \frac{|x-x'|^{4}}{4!} e^{-\frac{|x-x'|^{2}}{2a^{2}}}} + ... \Big)\Big] .
\end{split}
\end{equation*}
\end{widetext}

Evaluating the terms in the integral and cancelling the $\psi(y) \psi(z)$ terms from both sides, we obtain,

\begin{widetext}
\begin{equation}
\begin{split}
i\hbar\frac{\partial \psi(x,t)}{\partial t}&= - \frac{\hbar^{2}}{2m} \partial_{x}^{2}\psi(x,t)+ g \psi(x,t) \Big(|\psi(x,t)|^{2}\\
& +\frac{a^{2}}{2}\partial_{x}^{2}|\psi(x,t)|^{2} + \frac{a^{4}}{8}\partial_{x}^{4}|\psi(x,t)|^{2}+ \frac{a^{6}}{48} \partial_{x}^{6}|\psi(x,t)|^{2} + ... \Big) .
\label{eq:modifiedgp}
\end{split}
\end{equation}
\end{widetext}

The denominator of the numerical factors in the expansion are even double factrials and hence the numerical factors fall off. In our calculations of the amplitude modulated state, where $k$ is the wave number of the phase, we will see that $k\sim\frac{1}{a}$. Thus $ak$ is $O(1)$. Still it is safe to do the analysis here in a perturbative way, because the coefficients of the higher order terms would fall off rather quickly. We take the equation thus obtained as a modified GP equation and carry out our further analysis with the help of this equation retaining terms upto the $6th$ order. Since the coefficient of the next term would be an order of magnitude smaller than that of the $6th$ order term, we truncate the series at the decimal place corresponding to the $6th$ order term. This equation would capture the essential features of the effects of non-local interactions on the properties of a BEC. 
\par
A perturbative approach for the GP equation used, for example to determine the density profile of a single vortex line where the use of linear approximation for the density of the vortex core is made and subsequently numerical solution of the vortex density away from the core has been constructed, has given qualitative features of the vortex size and density profile. In spirit with this approximation, we propose that our perturbative approach too would be able to show the qualitative features of a density modulated state in a BEC.
\par
The first thing to note is that the uniform density solution $\psi_{0}=\sqrt{n}e^{-\frac{i\mu t}{\hbar}} $ is also a solution of the modified GP equation, where $\mu=gn$ and $|\psi_{0}|^{2}=n$ is the density of the condensate. Also note that we had to resort to a perturbative scheme here to study a pure density modulated state, since both equations (\ref{eq:original}) and (\ref{eq:contact}) do not admit a solution of the form $\cos{kx}/\sin{kx}$ even in one dimension as that would leave a cubic term in $\cos{kx}/\sin{kx}$ unbalanced. Hence, the Taylor expansion is a key ingredient in the recipe of obtaining a long range order as we shall see in the next section.

\section{Amplitude modulated phase}
 Let us consider a solution of Eq.(\ref{eq:modifiedgp}) of the form $\psi(x,t) = \psi (x) e^{\frac{-i\omega t}{\hbar}}$ where $\omega $ is the global oscillation frequency. $\omega$ gets identified as the chemical potential $\mu$ of the system in the case of a uniform ground state and would be of the same order for the modulated density state as we will see in the following. The uniform density GP ground state solution $\psi_0=\sqrt{n} e^{\frac{-i\mu t}{\hbar}} $ is still a solution of Eq.(\ref{eq:modifiedgp}) with the same free energy $F= \frac{gN^{2}}{2V}$ where the total number of particles $ N= n\int\vec{dr} = nV$ and $V$ is the volume. We would refer to this particular uniform density solution as the ground state frequently in what follows. There could be other single particle states as the solution of Eq.(\ref{eq:original}) as $\psi(\textbf{r},t) =\psi(x,t)\psi(y)\psi(z)=\sqrt{n} e^{i(kx -\frac{\omega t}{\hbar})}$ where $\psi(y)$ and $\psi(z)$ are unit constants. These are the solutions of the local GP equation as well, where there is a kinetic energy cost which makes them higher energy states as compared to the ground state at the same density. These are moving solutions with velocity $v=\hbar k/m$. The ground state is the $k \to 0$ limit of these single particle states.
\par
Eq.(\ref{eq:modifiedgp}) also admits solution where $ \psi(x) =A\cos{kx}$ or $ \psi(x) =A\sin{kx}$. Substituting $\cos{kx}$ or $\sin{kx}$ in Eq.(\ref{eq:modifiedgp}), we get $1-2a^{2}k^{2}+2a^{4}k^{4}-\frac{4a^{6}k^{6}}{3}= 0$ and $\omega =\frac{\hbar^{2}k^{2}}{2m}+g|A|^{2}(a^{2}k^{2}-a^{4}k^{4}+\frac{2a^{6}k^{6}}{3})$ , giving  $ k^{2} = 0.894 \left(\frac{1}{a^{2}}\right)$ and $\omega = \frac{\hbar^2k^2}{2m}+(0.57)g|A|^2$ ,where we have kept terms upto $a^{6}$ in the expansion.

\par
 The normalization condition over a length of $2L$ and a cross section $\sigma$ of the condensate gives

\begin{equation}
\frac{\sigma |A|^2}{2} \int_{-L}^L{dx(1\pm \cos{2kx})}=V|A|^2\left ( \frac{1}{2} \pm \frac{\sin{2kL}}{4kL}\right ) = N,
\end{equation}

where the upper sign of $\pm$ is for the $\cos{kx}$ profile and the lower one is for the $\sin{kx}$ profile (we will follow the same convention in what follows) and $V=2\sigma L$.

\par
\begin{figure}[h]
  \includegraphics[width=8cm]{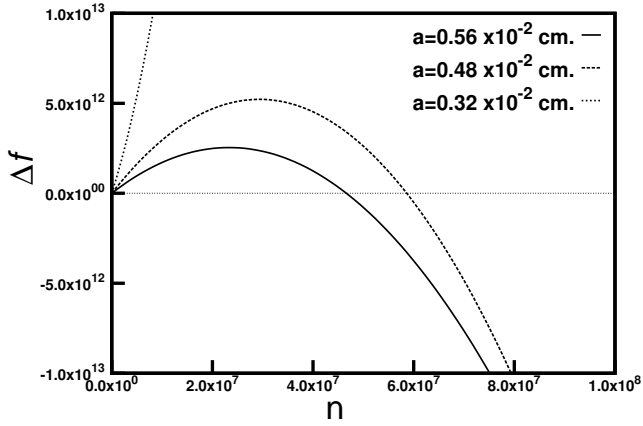}
  \caption{\small {Figure shows plot of $\Delta f$ versus the density($n$) for certain values of the s-wave scattering length $a$. L=0.02 cm.}}
\label{fig:intro_fig_1}
\end{figure}

Using the expression for free energy in Eq.(\ref{eq:energy}) we can compare the energies of the density modulated state($E_{m}$) and that of the uniform density state($E_{u}$). Fig.(\ref{fig:intro_fig_1}) shows the difference between the energy densities of the two states $\Delta f= E_{m} - E_{u}$ , for certain values of the s-wave scattering length $a$. Note here that $\Delta f$ going to negative values doesn't mean that the energy $E_{m}$ is negative. $E_{m}$ is always positive. $\Delta f$ becoming negative only means that as we change the density($n$), $E_{m}$ becomes less than $E_{u}$. In Fig.(\ref{fig:intro_fig_1}), the two for curves for which $\Delta f$ becomes negative, the difference $\Delta f$ between $E_{m}$ and $E_{u}$ is always less than $E_{u}=\frac{1}{2}gn^{2}$, meaning if we add $\frac{1}{2} gn^{2}$ to $\Delta f$, the resultant term would never go to zero, thus stressing on the point that $E_{m}$ is always positive.
\par
 In Fig.(\ref{fig:intro_fig_1}) , at lower density the energy of the density modulated state $A \cos{kx} e^{-i\omega t/\hbar}$ is greater than the uniform density state $\sqrt{n} e^{-\frac{i\mu t}{\hbar}}$ due to the kinetic energy term $\frac{-\hbar^{2}}{2m}\nabla^{2}\psi$ which is non zero for the modulated density state and zero for the uniform density state. The energy of the uniform density state comes only from interactions and hence to make the uniform density state to be of higher energy, we have to increase particle density which would make the interaction energy increase and eventually, as seen from Fig.(\ref{fig:intro_fig_1}), make the modulated density state energetically favourable under certain values of parameters (s-wave scattering length $a$ in Fig.(\ref{fig:intro_fig_1})). The crossover density is of the order of $n \sim \frac{k^2}{8\pi a}$ $\sim 10^{8} cm^{-3}$. So, the free energy difference decreases continuously beyond the above mentioned limit with the increase in density and, thus, fixing the density of the BEC at a suitable value one can expect to have such a state energetically favourable. This figure clearly shows that there are wide regions over which the free energy density of the ordered phase is less than that of the uniform ground state. Note here that although we are using a perturbative approach to find the $k$ selection for the density modulated state, we use the exact Free energy functional to find the energy of the density modulated state.

\par
If we take $L$ to be the length of the condensate and apply the periodic boundary conditions viz., $\psi(x=0)=\psi(x=L)$ i.e., $\cos{kL}=1$ , we get a selection on $k$ and hence on the scattering length $a$. We find that the values of $a$ that we get from the periodic boundary conditions cannot lower the energy of the density modulated state as compared to the uniform density state. To make the modulated energy state energetically favourable, $a$ has to be about one order of magnitude less than that of $L$ and this is where Feshbach resonance can come in handy. Since, the values of $a$ for which the density modulated state becomes energetically favourable does not obey periodic boundary conditions, it is clear that there would be an additional healing cost at the boundary. Since this healing cost is also present for the uniform density state hence the healing cost can be taken to be comparable for both these states. We envisage that by using feshbach resonance, probably it would be possible to make the boundary condition such that the energetically favourable amplitude modulated state shows up.
\par
Also, to note is the fact that here, thermodynamic limit does not make sense, because the values of the scattering length for which the density modulated state is energetically favourable is just $1$ order of magnitude smaller and hence hence both $L$ and $a$ would always be comparable.

\par
As already mentioned in the introduction, the density modulated state, like the uniform density state would be unstable since it does not rest in an energy minima, as shown schematically in Fig.(\ref{fig:schematic}). We show this by a linear stability analysis similar to that for the uniform density state, which goes as follows. To look at the stability of these states, let us consider the specific case $\psi(x,t)=A\cos{(kx)} e^{-i\omega t/\hbar}$ and perturb it by the small amplitude modes $\theta(x,t)=\sum_i{[u_i(x)e^{\frac{-i\omega_i t}{\hbar}}+v^*_i(x)e^{\frac{i\omega_i t}{\hbar}}]}e^{-i\omega t/\hbar}$. Taking the ansatz $u_i(x)=u_i e^{iqx}$ and $v_i(x)=v_i e^{iqx}$, the ensuing linear equations in the small amplitudes $u_i$ and $v_i$ would be of the form

\begin{widetext}
\begin{equation}
\begin{pmatrix} 
{(\alpha-\Phi)a^{2}A^{2}g\cos{kx}} & {\frac{-24\hbar^{2}q^{2}}{m}+\beta+(\alpha-\Phi)a^{2}A^{2}g\cos{kx}+48(\omega-\hbar\omega_{i})} \\ 
{\frac{-24\hbar^{2}q^{2}}{m}+\beta+(\alpha-\Phi^{*})a^{2}A^{2}g\cos{kx}+48(\omega+\hbar\omega_{i})} & {(\alpha-\Phi^{*})a^{2}A^{2}g\cos{kx}}
\end{pmatrix}
\begin{pmatrix}
u_i \\ v_i
\end{pmatrix}
= 0,
\end{equation}
\end{widetext}

where

\begin{widetext}
\begin{equation*}
\begin{split}
\alpha&= \Big(a^{4}k^{6}+3a^{2}k^{4}(-2+5a^{2}q^{2})+q^{2}(24-6a^{2}q^{2}+a^{4}q^{4})+3k^{2}(8-12a^{2}q^{2}+5a^{4}q^{4})\Big)\cos{kx},\\
\beta&= 16a^{2}A^{2}k^{2}(3-3a^{2}k^{2}+2a^{4}k^{4})\cos{2kx},\\
\Phi&= 2ikq\Big[24-12a^{2}(k^{2}+q^{2})+a^{4}(3k^{4}+10k^{2}q^{2}+3q^{4})\Big] \sin{kx}.
\end{split}
\end{equation*}
\end{widetext}
\par
Because of the presence of the irremovable imaginary term in the expression of $\Phi$, $\omega_i$ is always complex and this indicates an instability of the ordered phase arising out of the coupling of the small excitations to the amplitude modulated state in the correction term representing the non-local interactions. Thus, within the scope of the perturbative approach adopted here to obtain an explicit amplitude modulated solution of BEC using s-wave scattering, the amplitude modulated state is generically dynamically unstable but probably could be seen on a short lived state in a BEC by appropriately tuning Feshbach resonance.

\section{Discussions}
We propose a new approach to obtain pure density modulated state as a zero temperature state of the system by looking directly at the long range interactions between particles of the BEC and considering their effects on the order parameter. This approach is very different from the usual approach of looking at small amplitude excitations on top of the uniform density state and lowering the roton minimum. By using Taylor expansion technique, we were able to obtain a density modulated state as a perturbative solution to GP equation with long range order. Perturbative approach had to be used as the conventional GP equation does not admit modulated density solutions. Although, the solution is perturbative, we use the exact free energy functional to evaluate energy. Thus using this sort of perturbative scheme starting from the GP equation is the key idea of this article.
\par
By energy calculations, we have shown that there exist range of the scattering length $a$ where this density modulated state becomes energetically favourable as compared to the uniform density state. This range of $a$ depends on the length of the condensate and is an order of magnitude smaller than the condensate length. The crossover density we have obtained is of the order of $10^{8} cm^{-3}$. This shows that the density modulations would manifest themselves even when the density is as low as $10^{8}$ - $10^{10} cm^{-3}$. Also, we have seen that the $a$ selection obtained by the periodic boundary conditions does not enable making the density modulated state energetically favourable, meaning that the energy lowering as compared to the uniform density state, comes from the boundaries. Since, there is only a specific range of $a$ for which this energy lowering is possible, Feshbach resonance can play a key role in obtaining a density modulated state.

\par
We have shown that as such density modulated state, even though energetically favourable than the uniform density state, isn't stable under small amplitude oscillations as it does not sit in an energy minima. This instability is present in the case of the uniform density state too. Thus, it is clear that we have to go beyond the s-wave interactions in hope of stabilizing such a modulated density state. In this regard, we would look to dipolar interactions within the perturbativbe scheme that we have proposed, in the future. We feel that by looking at the perturbative scheme that we have proposed, it might be possible to make many more predictions about the various aspects of BEC.

\end{document}